\documentclass[runningheads]{llncs}
\usepackage{graphicx}
\usepackage[dvipsnames]{xcolor}
\usepackage{amsmath}
\usepackage{amsfonts}
\usepackage{caption}
\usepackage{subcaption}
\usepackage{microtype}
\usepackage{cite}
\usepackage{booktabs}
\usepackage[colorlinks = true, linkcolor = blue, citecolor = blue]{hyperref}

\begin{document}
\title{Deep filter bank regression for super-resolution of anisotropic MR brain images}
\titlerunning{Deep filter bank regression}

\author{
    Samuel~W.~Remedios\inst{1}, 
    Shuo~Han\inst{2},
    Yuan~Xue\inst{3},
    Aaron~Carass\inst{3},\\
    Trac~D.~Tran\inst{3},
    Dzung~L.~Pham\inst{4}, and
    Jerry~L.~Prince\inst{3}
}

\authorrunning{S.W. Remedios et al.}

\institute{
    $^1$~Department of Computer Science,\\
    Johns Hopkins University, Baltimore,~MD~21218,~USA\\
    \email{samuel.remedios@jhu.edu}\\[0.75em]
    $^2$~Department of Biomedical Engineering,\\
    Johns Hopkins University School of Medicine, Baltimore,~MD~21205,~USA\\[0.75em]
    $^3$~Department of Electrical and Computer Engineering,\\Johns Hopkins University, Baltimore,~MD~21218,~USA\\[0.75em]
    $^4$~Center for Neuroscience and Regenerative Medicine, Henry~M.~Jackson~Foundation,~Bethesda,~MD~20817,~USA
}

\maketitle              

\begin{abstract}
In 2D multi-slice magnetic resonance (MR) acquisition, the through-plane signals are typically of lower resolution than the in-plane signals. While contemporary super-resolution~(SR) methods aim to recover the underlying high-resolution volume, the estimated high-frequency information is implicit via end-to-end data-driven training rather than being explicitly stated and sought. To address this, we reframe the SR problem statement in terms of perfect reconstruction filter banks, enabling us to identify and directly estimate the missing information. In this work, we propose a two-stage approach to approximate the completion of a perfect reconstruction filter bank corresponding to the anisotropic acquisition of a particular scan. In stage~1, we estimate the missing filters using gradient descent and in stage~2, we use deep networks to learn the mapping from coarse coefficients to detail coefficients. In addition, the proposed formulation does not rely on external training data, circumventing the need for domain shift correction. Under our approach, SR performance is improved particularly in ``slice gap'' scenarios, likely due to the constrained solution space imposed by the framework. 

\end{abstract}
\section{Introduction}
Anisotropic magnetic resonance (MR) images are those acquired with high in-plane resolution and low through-plane resolution. It is common practice to acquire anisotropic volumes in clinics as it reduces scan time and motion artifacts while preserving signal-to-noise ratio. To improve through-plane resolution, data-driven super-resolution~(SR) methods have been developed on MR volumes~\cite{zhao2020smore,oktay2016multi,chen2018efficient,du2020super}. The application of SR methods to estimate the underlying isotropic volume has been shown to improve performance on downstream tasks~\cite{zhao2019applications}.

For 2D multi-slice protocols, the through-plane point-spread function~(PSF) is known as the slice profile. When the sampling step is an integer, the through-plane signals of an acquired MR image can be modeled as a strided 1D convolution between the slice profile and the object to be imaged~\cite{han2021mr,prince2006medical,sonderby2016amortised}. Commonly, the separation between slices is equivalent to the full-width-at-half-max~(FWHM) of the slice profile, but volumes can also be acquired where the slice separation is less than or greater than the slice profile FWHM, corresponding to ``slice overlap'' and ``slice gap'' scenarios, respectively.

Data-driven SR methods usually simulate low-resolution~(LR) data from high-resolution~(HR) data using an assumed slice profile~\cite{zhao2020smore,oktay2016multi,chen2018efficient,du2020super}, or an estimated slice profile according to the image data or acquisition~\cite{han2021mr}. In either case, SR methods are generally formulated as a classical inverse problem:
\begin{equation}
    y = Ax,
\label{eq:inverse}
\end{equation}
where $y$ is the LR observation, $A$ is the degradation matrix, and $x$ is the underlying HR image. Commonly, this is precisely how paired training data is created for supervised machine learning methods; HR data is degraded by $A$ to obtain the LR $y$ and weights $\theta$ of a parameterized function $\phi$ (e.g., a neural network) are learned such that $\phi_\theta(y) \approx x$. However, under this framework there is no specification of information lost by application of $A$; contemporary SR models train end-to-end and are directed only by the dataset.

\begin{figure}[!tb]
    \centering
    \includegraphics[width=\textwidth]{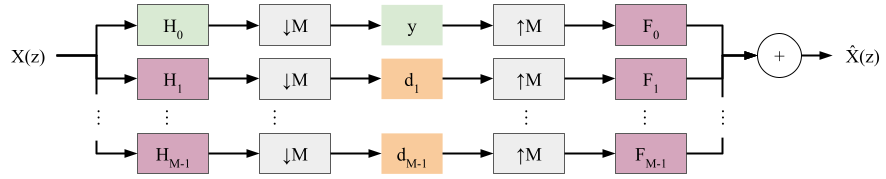}
    \caption{The filter bank observation model. Both $y$ and $H_0$ (green) are given and fixed. In stage 1, filters $H_1, \ldots, H_{M-1}$ and $F_0, F_1, \ldots, F_{M-1}$ (purple) are learned; in stage 2, a mapping from $y$ to $d_1, \ldots, d_{M-1}$ (gold) is learned.}
    \label{fig:obs_model}
\end{figure}

In our work, we propose an entirely novel SR framework based on perfect reconstruction~(PR) filter banks. From filter bank theory, PR of a signal $x$ is possible through an $M$-channel filter bank with a correct design of an analysis bank $H$ and synthesis bank $F$~\cite{strang_nguyen_1997}. Under this formulation, we do not change Eq.~\ref{eq:inverse} but explicitly recognize our observation $y$ as the ``coarse approximation'' filter bank coefficients and the missing information necessary to recover $x$ as the ``detail'' coefficients (see Fig.~\ref{fig:obs_model}). For reference, in machine learning jargon, the analysis bank is an encoder, the synthesis bank is a decoder, and the coarse approximation and detail coefficients are analogous to a ``latent space.''

The primary contribution of this work is to reformulate SR to isotropy of 2D-acquired MR volumes as a filter bank regression framework. The proposed framework has several benefits. First, the observed low-frequency information is untouched in the reconstruction; thus, our method explicitly synthesizes the missing high frequencies and does not need to learn to preserve acquired low frequency information. Second, the downsampling factor $M$ specifies the number of channels in the $M$-channel filter bank, constraining the solution space in tougher scenarios such as ``slice gap'' acquisition recovery. Third, the analysis filters of PR filter banks necessarily introduce aliasing which is canceled via the synthesis filters; therefore, we do not need to directly handle the anti-aliasing of the observed image. Fourth, our architecture has a dynamic capacity for lower-resolution images. The rationale behind the dynamic capacity is intuitive: when fewer measurements are taken, more estimates must be done in recovery and a more robust model is required. Fifth, our method exploits the nature of anisotropic volumetric data; in-plane slices are HR while through-plane slices are LR. Thus, we do not rely on external training data and only need the in-plane HR data to perform internal supervision. In the remainder of the paper, we describe this framework in detail, provide implementation details, and evaluate against a state-of-the-art internally supervised SR technique. We demonstrate the feasibility of formulating SR as filter bank coefficient regression and believe it lays the foundation for future theoretical and experimental work in SR of MR images.

\section{Methods}
%
%
The analysis bank $H$ and synthesis bank $F$ each consist of $M$ 1D filters represented in the $z$-domain as $H_k$ and $F_k$, respectively, with corresponding spatial domain representations $h_k$ and $f_k$.
As illustrated in Fig.~\ref{fig:obs_model}, input signal $X(z) = \mathcal{Z}(x)$\footnote{$\mathcal{Z}(x)$ is the $\mathcal{Z}$-transform of $x$\cite{strang_nguyen_1997}.} is filtered by $H_k$, then decimated with $\downarrow M$ (keeping every $M^\text{th}$ entry) to produce the corresponding coefficients. These coefficients exhibit aliasing and distortion which are corrected by the synthesis filters~\cite{strang_nguyen_1997}. Reconstruction from coefficients comes from zero-insertion upsampling with $\uparrow M$, passing through filters $F_k$, then summing across the $M$ channels.

\begin{figure}[!tb]
    \centering
    \includegraphics[
        width=0.9\textwidth, 
        page=2,
        trim=3cm 37cm 25cm 7cm, 
        clip,
    ]{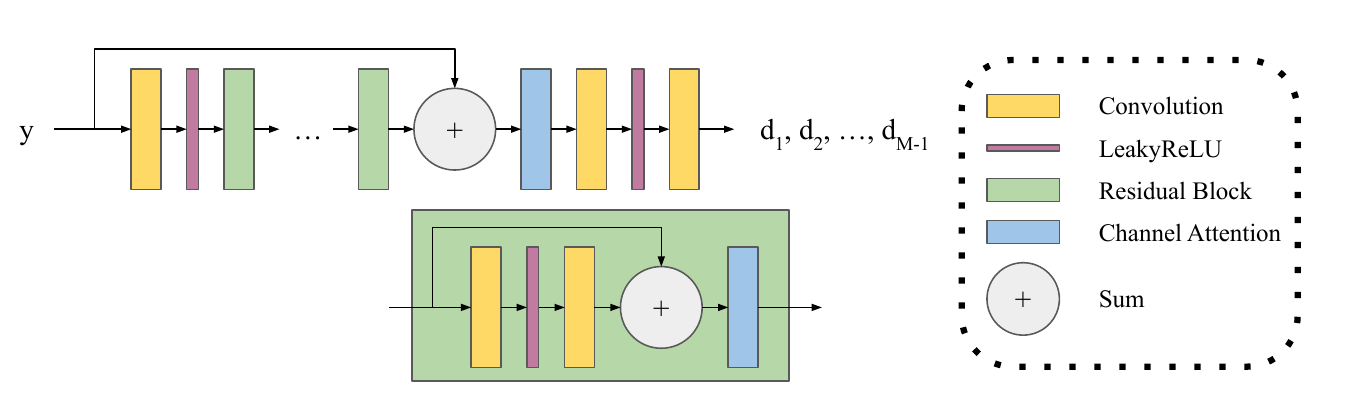}
    \caption{This network architecture, used in the second stage of our algorithm, has the same structure for both the generator and discriminator but with different hyperparameters. All convolutional layers used a $3\times 3$ kernel. The generator and discriminator used $16$ and $2$ residual blocks, respectively. The generator had $128 \times M$ features per convolutional layer while the discriminator had $64 \times M$ features per convolutional layer. The final convolution outputs $M-1$ channels corresponding to the missing filter bank detail coefficients. The internal structure of the residual block is encapsulated in green.}
    \label{fig:network}
\end{figure}

Traditional design of $M$-channel PR filter banks involves a deliberate choice of a prototype low-pass filter $H_0$ such that modulations and alternations of the prototype produce the remaining filters for both the analysis and synthesis filter banks~\cite{strang_nguyen_1997}. $M$ is also chosen based on the restrictions of the problem at hand. However, for anisotropic 2D-acquired MRI, the slice profile \textit{is} the low-pass filter and as such we have a fixed, given $H_0$. The separation between slices is equal to the FWHM of $h_0$ plus any further gap between slices. We denote the slice separation as $M$, corresponding to the number of channels in the PR filter bank. We use $A \| B$, read ``A skip B'', to denote a FWHM of $A$~mm and slice gap of $B$~mm and note that $M = A + B$. For this preliminary work, we assume $A, B$, and $M$ are all integer and, without loss of generality, assume that the in-plane resolution is $1 \| 0$.

Our goal is to estimate filters $H_1, \ldots, H_{M-1}$ and $F_0, \ldots, F_{M-1}$ and the detail coefficients $d_1, \ldots, d_{M-1}$ which lead to PR of $x$. We approach this problem in two stages. In stage~1, we approximate the missing analysis and synthesis filters, assuming there exists a set of filters to complete the $M$-channel PR filter bank given that $H_0$ and $M$ are fixed and known ahead of time. These must be learned first to establish the approximate PR filter bank conditions on the coefficient space. Then, in stage~2, we perform a regression on the missing coefficients. Both of these stages are optimized in a data-driven end-to-end fashion with gradient descent. After training, our method is applied by regressing $d_1, \ldots, d_{M-1}$ from $y$ and feeding all coefficients through the synthesis bank to produce $\hat{x}$, our estimate of the HR signal. The stage~2 coefficient regression occurs in 2D, so we construct our estimate of the 3D volume by averaging stacked 2D predictions from the synthesis bank from both cardinal planes containing the through-plane axis.

\textbf{Stage 1: Filter Optimization}\qquad{}Previous works assumed the slice profile is Gaussian with FWHM equal to the slice separation~\cite{zhao2020smore,oktay2016multi}; instead, we estimate the slice profile, $H_0$, directly with ESPRESO\footnote{\url{https://github.com/shuohan/espreso2}}~\cite{han2021mr}. We next aim to estimate the filters $H_1, \ldots, H_{M-1}$ and $F_0, \ldots, F_{M-1}$. To achieve this, we learn the spatial representations $h_1, \ldots, h_{M-1}$ and $f_0, \ldots, f_{M-1}$ from 1D rows and columns drawn from the high resolution in-plane slices of $y$, denoted $\mathcal{D}_1 = \{x_i\}_{i=1}^N$. We initialize these filters according to a cosine modulation~\cite{strang_nguyen_1997} of $h_0$, which is defined as 
\begin{equation*}
    f_k[n] = h_k[n] = h_0[n] \sqrt{\frac{2}{M}} \cos{\left[
         \left( k + \frac{1}{2} \right)
         \left( n + \frac{M + 1}{2} \right)
         \frac{\pi}{M}\right]
        },
\end{equation*} for $ k \in \{1, 2, \ldots, M-1\}$.
Accordingly, we initialize $f_0$ to $h_0$. We estimate $\hat{x}_{i}$ by passing $x_i$ through the analysis and synthesis banks, then (since the entire operation is differentiable) step all filters except $h_0$ through gradient descent. The reconstruction error is measured with mean squared error loss and the filters are updated based on the AdamW~\cite{loshchilov2017decoupled} optimizer with a learning rate of $0.1$, the one-cycle learning rate scheduler~\cite{smith2019super}, and a batch size of $32$ for $100,000$ steps.

\textbf{Stage 2: Coefficient Regression}\qquad{}From stage~1, we have the analysis and synthesis banks and now want to estimate the missing detail coefficients given only the LR observation $y$. With the correct coefficients and synthesis filters, PR of $x$ is possible. For this stage, we use 2D patches, in spite of the 1D SR problem, as a type of ``neighborhood regularization''. Let $\mathcal{D}_2 = \{x_i\}_{i=1}^N$, $x_i \in \mathbb{R}^{p \times pM}$; i.e., the training set for stage~2 consists of 2D $p \times pM$ patches drawn from the in-plane slices of $y$. The second dimension will be decimated by $M$ after passing through the analysis banks, resulting in $y, d_1, \ldots, d_{M-1} \in \mathbb{R}^{p \times p}$.  We use the analysis bank (learned in stage~1) to create training pairs $\{(y_i, (d_1, d_2, \ldots, d_{M-1})_i)\}_{i=1}^N$ and fit a convolutional neural network (CNN) $G: \mathbb{R}^{p \times p} \rightarrow \mathbb{R}^{{p \times p}^{M-1}}$ to map $y_i$ to $(d_1, \ldots, d_{M-1})_i$. In this work, we set $p = 32$. Since this is an image-to-image translation task, we adopt the widely used approach proposed in Pix2Pix~\cite{pix2pix2017} including the adversarial patch discriminator. 

Empirically, we found more learnable parameters are needed with greater $M$. Thus, our generator $G$ is a CNN illustrated in Fig.~\ref{fig:network} with $16$ residual blocks and $128 \times M$ kernels of size $3\times 3$ per convolutional layer. The discriminator $D$ has the same architecture but with only $2$ residual blocks and $64 \times M$ kernels per convolutional layer. Our final loss function for stage~2 is identical to the loss proposed in~\cite{pix2pix2017} and is calculated on the error in $(d_1, \ldots, d_{M-1})_i$. We use the AdamW optimizer~\cite{loshchilov2017decoupled} with a learning rate of $10^{-4}$ and the one-cycle learning rate scheduler~\cite{smith2019super} for $500,000$ steps at a batch size of $32$.

\section{Experiments and Results}

\begin{table}[!tb]
    \centering
    \caption{Mean $\pm$ std. dev. of volumetric PSNR values for stage~1 reconstruction of the $30$ subjects. ``Self'' indicates a reconstruction of the input low-resolution volume on which the filter bank was optimized, while ``GT'' indicates reconstruction of the isotropic ground truth volume. (L-R), (A-P), and (S-I) are the left-to-right, anterior-to-posterior, and superior-to-inferior directions, respectively.}
    \label{tab:autoencoding}
    \begin{tabular}{c|c|c|c|c|c}
    \toprule
    \hspace*{9ex}
    & Self (L-R)
    & Self (A-P)
    & GT (L-R)
    & GT (A-P)
    & GT (S-I)\\
    
    \cmidrule{1-6}
        $2\|0$& ~$62.24\pm 0.97$~ & ~$60.19\pm 3.74$~ & ~$60.63\pm 0.56$~ & ~$59.59\pm 2.54$~ & ~$55.47\pm 4.69$\\
        
        $2\|1$ & $63.01\pm 4.91$ & $62.25\pm 5.09$ & $64.32\pm 0.63$ & $59.49\pm 5.52$ & $53.81\pm 6.50$\\
    
        $2\|2$ & $62.57\pm 1.59$ & $57.93\pm 5.32$ & $60.62\pm 1.34$ & $59.31\pm 3.65$ & $52.09\pm 4.34$\\
        
    \cmidrule{1-6}
        $4\|0$ & $55.47\pm 3.81$ & $52.36\pm 5.32$ & $48.91\pm 4.65$ & $48.77\pm 4.68$ & $44.08\pm 4.78$\\
    
        $4\|1$ & $53.03\pm 1.54$ & $50.31\pm 3.41$ & $44.19\pm 1.57$ & $45.65\pm 1.63$ & $44.28\pm 2.14$\\
    
        $4\|2$ & $54.71\pm 2.61$ & $51.08\pm 4.51$ & $46.75\pm 2.83$ & $46.39\pm 3.27$ & $43.27\pm 2.80$\\
        
    \cmidrule{1-6}
        $6\|0$ & $49.97\pm 1.07$ & $40.18\pm 4.77$ & $40.14\pm 1.35$ & $41.04\pm 1.40$ & $35.76\pm 3.19$\\
    
        $6\|1$ & $52.35\pm 0.55$ & $45.69\pm 5.24$ & $42.11\pm 0.84$ & $42.74\pm 1.25$ & $39.76\pm 3.47$\\
    
        $6\|2$ & $53.17\pm 3.17$ & $49.11\pm 3.41$ & $43.66\pm 4.12$ & $44.87\pm 3.99$ & $41.50\pm 2.29$\\
        
    \bottomrule
    \end{tabular}
\end{table}

\begin{figure}[!tb]
    \centering
    \includegraphics[width=\textwidth]{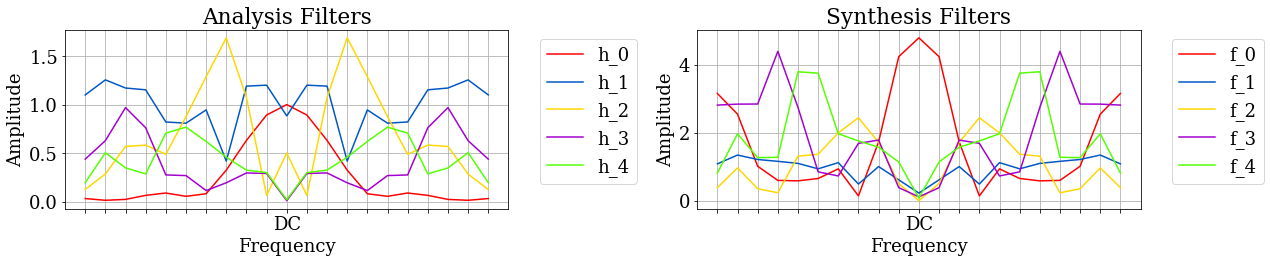}
    \caption{Estimated PR filters from stage~1 for a single subject at $4\| 1$ resolution ($M = 5$) in the frequency domain. Note the amplitudes for analysis and synthesis banks are on different scales, DC is centered, and $h_0$ is estimated by ESPRESO~\cite{han2021mr}.}
    \label{fig:example_filters}
\end{figure}

\begin{figure}[!tbp]
    \centering
    \includegraphics[
        width=\textwidth, 
        page=1,
        trim=4cm 0cm 7.5cm 0cm, 
        clip,
    ]{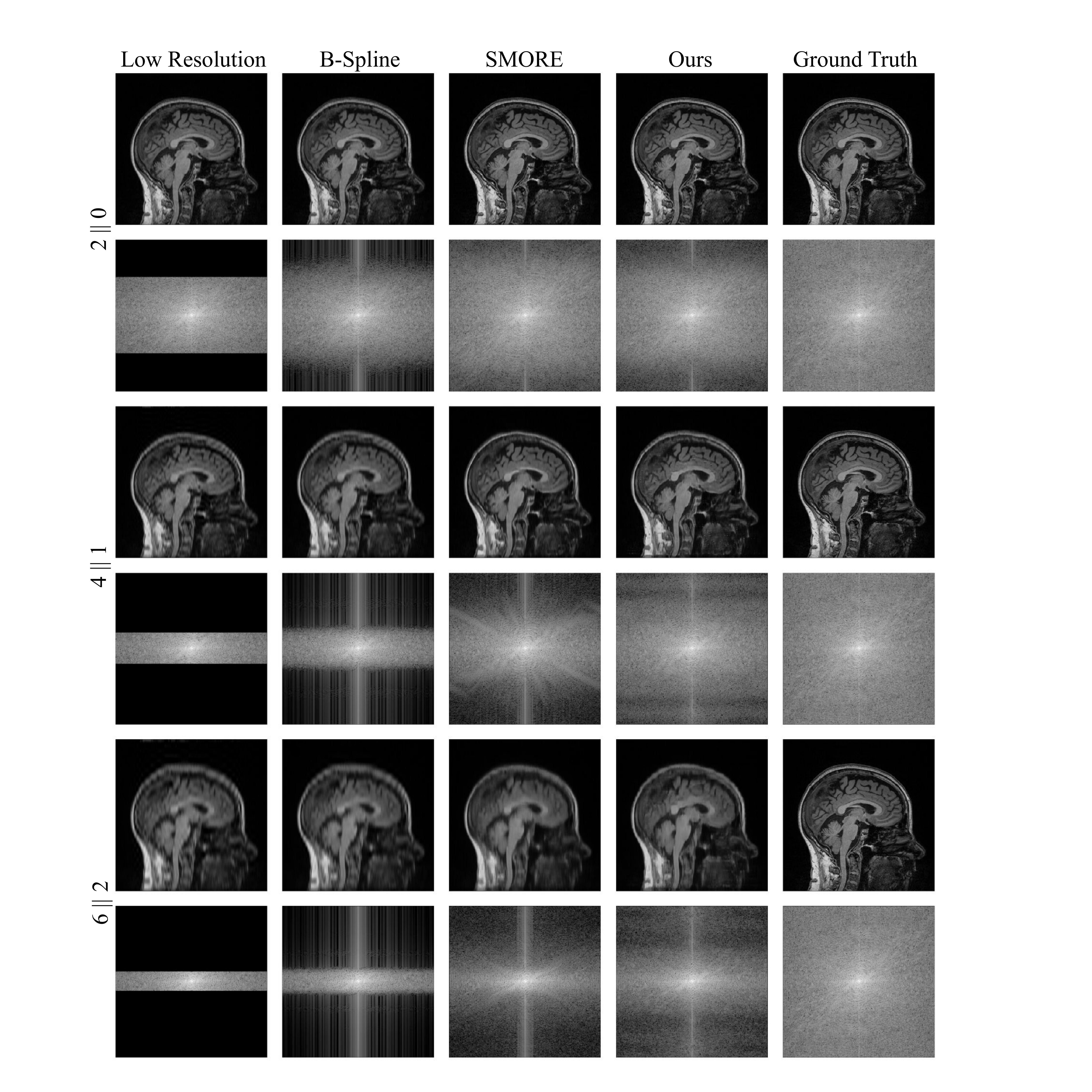}
    \caption{Mid-sagittal slice for a representative subject at different resolutions and gaps for each method. The low resolution column is digitally upsampled with k-space zero-filling. $A\| B$ signifies a slice thickness of $A$~mm and a gap of $B$~mm. Fourier magnitude is displayed in dB on every other row. The top two rows correspond to $2\| 0$ ($M=2$) for the MR slice and Fourier space, the second two rows are for $4\| 1$ ($M=5$), and the bottom two rows are for $6\| 2$ ($M=8$).
    }
    \label{fig:qualitative}
\end{figure}

\begin{figure}[!tb]
    \centering
    \begin{tabular}{c c c}
    \includegraphics[width=0.47\textwidth]{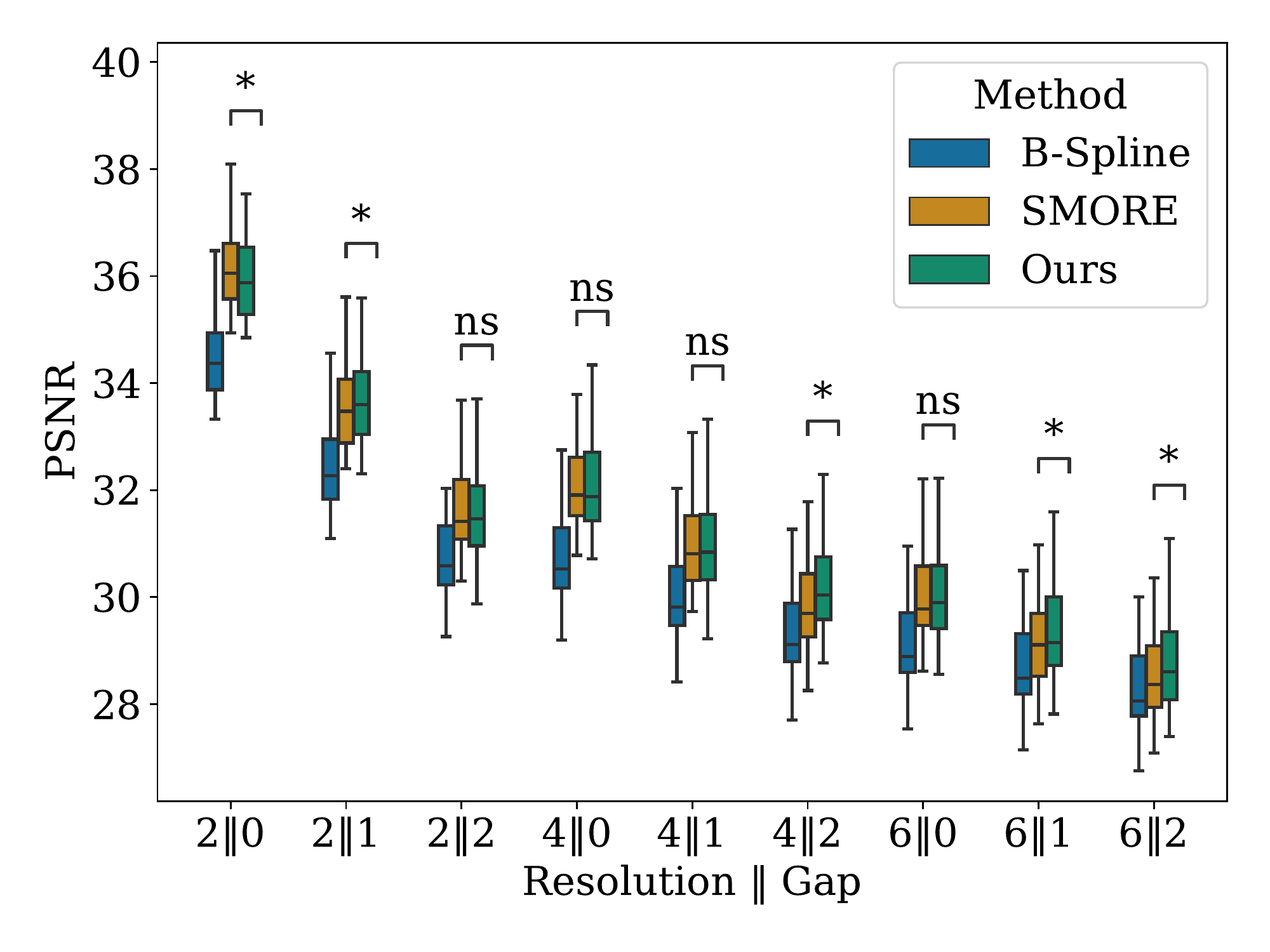} && \includegraphics[width=0.47\textwidth]{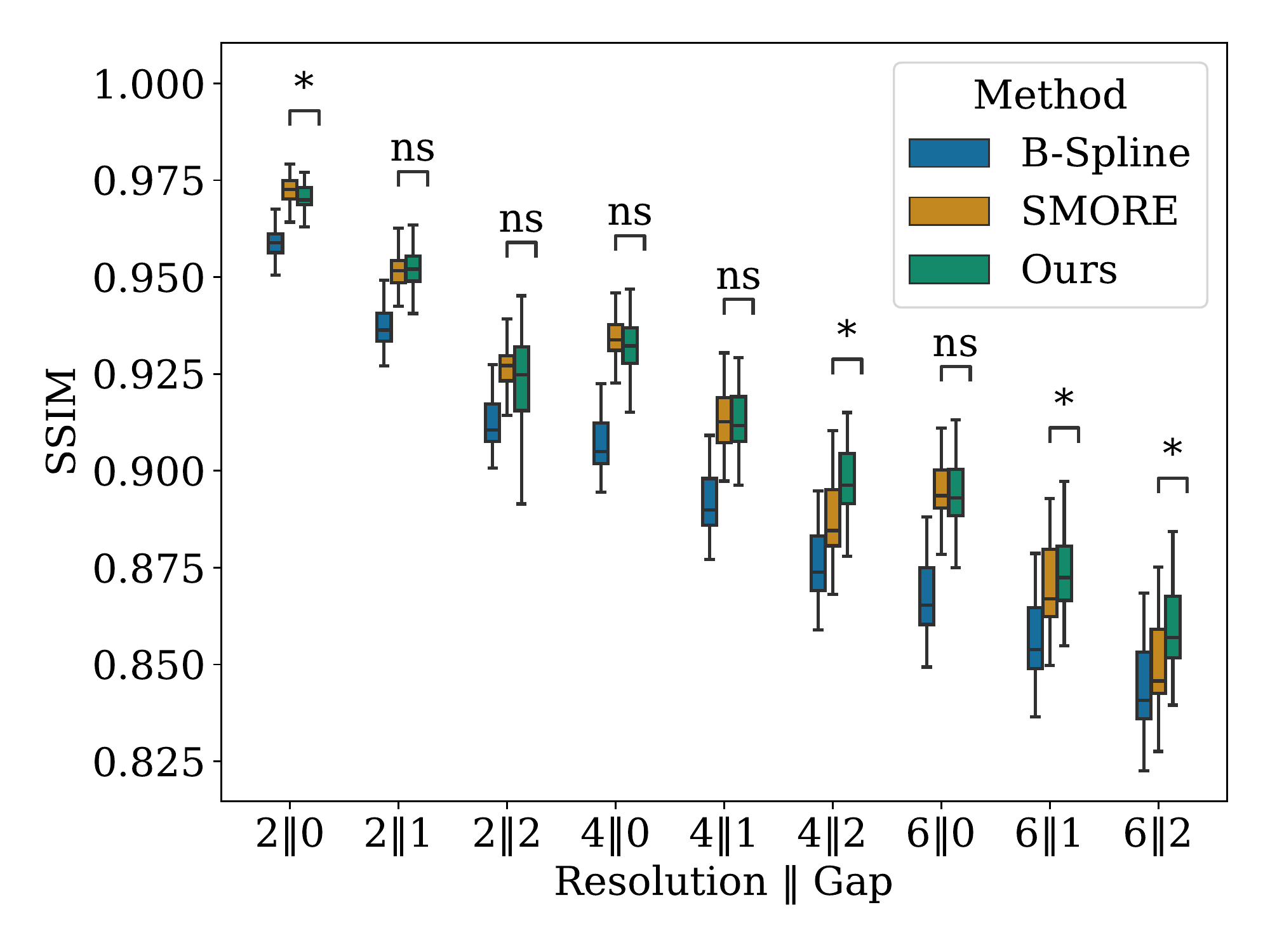}
    \\[-0.5em]
    \textbf{(a)} && \textbf{(b)}\\[0em]
    \end{tabular}
    \caption{Quantitative metrics PSNR in \textbf{(a)} and SSIM in \textbf{(b)}, computed over the $30$ image volumes. Significance tests are performed between SMORE and our proposed method with the Wilcoxon signed rank test; $\ast$ denotes $p$-values $ < 0.05$; ``ns'' stands for ``not significant''.}
    \label{f:oasis30}
\end{figure}

\textbf{Experiments}\qquad{}We performed two experiments to evaluate the efficacy of each stage in our approach. We randomly selected $30$ T1-weighted MR brain volumes from the OASIS-3 dataset~\cite{LaMontagne2019_OASIS3} to validate both stages and simulated LR acquisition via convolution with a Gaussian kernel with FWHM $\in \{2, 4, 6\}$ and slice gap $\in \{0, 1, 2\}$, yielding nine combinations of FHWM and slice gap in total. Since $M = A + B$ for a scan of resolution $A \| B$,  $M \in \{2, 3, 4, 5, 6, 7, 8\}$. For these experiments, the HR plane was axial while the cardinal LR planes were sagittal and coronal. We note that both stage~1 and stage~2 are trained for each LR volume separately as our proposed method does not use external training data, but instead relies on the inherent anisotropy in the multi-slice volume (i.e., HR in-plane and LR through-plane data).

\textbf{Stage 1 Results}\qquad{}We trained stage~1 using both cardinal 1D directions from in-plane data; that is, left-to-right (L-R) and anterior-to-posterior (A-P) directions. We then performed 1D reconstruction along these cardinal directions and collated all reconstructions into 3D volumes. In other words, this is an evaluation of self-auto-encoding. The mean volumetric reconstruction PSNR $\pm$ std. dev. across the $30$ subjects is shown in Table~\ref{tab:autoencoding}. In addition to applying the learned filters to the LR image itself, we would also like to test the extent of signal recovery for the HR counterpart that is the ground truth (GT) of the LR volume. Indeed, the coefficients generated by our learned analysis bank are what we will regress in stage~2, so a reconstruction of the GT is also shown in the right three columns of Table~\ref{tab:autoencoding}. This serves as a sort of ``upper bound'' on our super-resolution estimate and also answers the question of how well internal training generalizes to reconstruction of an isotropic volume.

We note that if we had attained PR filters, the PSNR would be $\infty$; our estimates fall short of this. Notably, reconstruction performance drops in the (S-I) direction; this is likely due to the fact that signals along this direction were not included in the training data. Additionally, an example of learned filters in the frequency domain for one resolution, $4\| 1$ ($M=5$), is shown in Fig.~\ref{fig:example_filters}. Recall that the fixed filter $h_0$ is the slice selection profile. We observe that our optimization approximated bandpass filters. 

\textbf{Stage 2 Results}\qquad{}To evaluate stage~2, we compared our method to two approaches which also do not rely on external training data: cubic b-spline interpolation and SMORE~\cite{zhao2020smore}, a state-of-the-art self-super-resolution technique for anisotropic MR volumes. For a fair comparison and improving SMORE results, SMORE was trained with the same slice profile that we use (the ESPRESO estimate~\cite{han2021mr}) instead of a Gaussian slice profile used in the original paper.

Qualitative results are displayed in Fig.~\ref{fig:qualitative} of a mid-sagittal slice for a representative subject at $2\| 0$, $4\| 1$, and $6\| 2$. This subject is near the median PSNR value for that resolution across the $30$ subjects evaluated in our experiments and for which SMORE outperforms our method at $2\| 0$, is on par with our method at $4\| 1$, and is outperformed by our method at $6\| 2$. Also shown in Fig.~\ref{fig:qualitative} is the corresponding Fourier space, and we see that our proposed method includes more high frequencies than the other methods. For quantitative results, PSNR and SSIM were calculated on entire volumes, as illustrated in box plots in Fig.~\ref{f:oasis30}.   



%
%

\section{Discussion and conclusions}
In this paper, we have presented a novel filter bank formulation for SR of 2D-acquired anisotropic MR volumes as the regression of filter-specified missing detail coefficients in an $M$-channel PR filter bank that does not change the low-frequency sub-bands of the acquired image. We would emphasize that our approach establishes a new theoretic basis for SR. In theory, these coefficients exist and give exact recovery of the underlying HR signal. However, it is unknown whether a mapping of $y \rightarrow (d_1, \ldots, d_{M-1})$ exists, and whether it is possible to find filters to complete the analysis and synthesis banks to guarantee PR. In practice, we estimate these in two stages: stage~1 estimates the missing analysis and synthesis filters towards PR and stage~2 trains a CNN to regress the missing detail coefficients given the coarse approximation $y$. According to our experiments, as the resolution worsens and slice gap increases our proposed method better handles the SR task than the competitive approach, validating the usefulness of our method for super resolving anisotropic MR images with large slice gaps. Future work will include: 1)~deeper investigation into the limits of the training set in learning the regression; 2)~the degree to which the mapping $G$ is valid; 3)~more analysis of the super-resolved frequency space; and 4)~develop methods to exactly achieve or better approximate PR. True PR filter banks should greatly improve the method, as Table~\ref{tab:autoencoding} serves as a type of ``upper bound'' for our method; regardless of the quality of coefficient regression, even given the ideal ground truth coefficients, reconstruction accuracy would be limited. Furthermore, our work suffers two major shortcomings. First, our current assumptions are integer slice thickness and slice separation, which is not always true in reality. To address this, the use of fractional sampling rates with filter banks~\cite{strang_nguyen_1997} may be a promising research direction. Second, our model in stage 2 scales the number of convolution kernels per layer by $M$. This induces a longer training and testing time when the image is of lower resolution. For reference, SMORE produced the SR volume in about 86 minutes on a single NVIDIA V100 regardless of the input resolution, but our proposed method produced the SR volume in 27 minutes for $2 \| 0$, 85 minutes for $4 \| 1$, and 127 minutes for $6 \| 2$.  Additionally, further investigation into improved regression is needed---a model which can better capture the necessary aliasing in the coefficient domain is vital for PR.

\section{Acknowledgements}
This material is supported by the National Science Foundation Graduate Research Fellowship under Grant No. DGE-1746891. Theoretical development is partially supported by NIH ORIP grant R21~OD030163 and the Congressionally Directed Medical Research Programs (CDMRP) grant MS190131. This work also received support from National Multiple Sclerosis Society RG-1907-34570, CDMRP W81XWH2010912, and the Department of Defense in the Center for Neuroscience and Regenerative Medicine.

\bibliographystyle{splncs04}
\bibliography{refs}
\end{document}